\documentclass[letterpaper]{article} % DO NOT CHANGE THIS
\usepackage{aaai2026}  % DO NOT CHANGE THIS
\usepackage{times}  % DO NOT CHANGE THIS
\usepackage{helvet}  % DO NOT CHANGE THIS
\usepackage{courier}  % DO NOT CHANGE THIS
\usepackage[hyphens]{url}  % DO NOT CHANGE THIS
\usepackage{graphicx} % DO NOT CHANGE THIS
\usepackage{natbib}  % DO NOT CHANGE THIS AND DO NOT ADD ANY OPTIONS TO IT
\usepackage{caption} % DO NOT CHANGE THIS AND DO NOT ADD ANY OPTIONS TO IT
\usepackage{algorithm}
\usepackage{algorithmic}

\usepackage{enumitem}
\usepackage{amsmath}
\usepackage{multirow}
\usepackage{booktabs}
\usepackage{makecell}

\usepackage{newfloat}
\usepackage{listings}
\DeclareCaptionStyle{ruled}{labelfont=normalfont,labelsep=colon,strut=off} % DO NOT CHANGE THIS
\floatstyle{ruled}
\newfloat{listing}{tb}{lst}{}
\floatname{listing}{Listing}
\title{SpeakerLM: End-to-End Versatile Speaker Diarization and Recognition with Multimodal Large Language Models}
\author{
    Han Yin\equalcontrib,
    Yafeng Chen\equalcontrib,
    Chong Deng,
    Luyao Cheng,
    Hui Wang, \\
    Chao-Hong Tan,
    Qian Chen,
    Wen Wang, 
    Xiangang Li
}
\affiliations{
    \textsuperscript{\rm 1}Tongyi Fun Team, Alibaba Group\\
    yinh72906@gmail.com, yfchen97@mail.ustc.edu.cn
}

\usepackage{bibentry}
\begin{document}

\maketitle

\begin{abstract}
The Speaker Diarization and Recognition (SDR) task aims to predict ``who spoke when and what'' within an audio clip, which is a crucial task in various real-world multi-speaker scenarios such as meeting transcription and dialogue systems. Existing SDR systems typically adopt a cascaded framework, combining multiple modules such as speaker diarization (SD) and automatic speech recognition (ASR). The cascaded systems suffer from several limitations, such as error propagation, difficulty in handling overlapping speech, and lack of joint optimization for exploring the synergy between SD and ASR tasks. To address these limitations, we introduce \textbf{SpeakerLM}, a \textbf{unified multimodal large language model for SDR} that jointly performs SD and ASR in an \textbf{end-to-end} manner. Moreover, to facilitate diverse real-world scenarios, we incorporate a \textbf{flexible speaker registration mechanism} into SpeakerLM, enabling SDR under different speaker registration settings. SpeakerLM is progressively developed with a multi-stage training strategy on large-scale real data. Extensive experiments show that SpeakerLM demonstrates strong data scaling capability and generalizability, outperforming state-of-the-art cascaded baselines on both in-domain and out-of-domain public SDR benchmarks. Furthermore, experimental results show that the proposed speaker registration mechanism effectively ensures robust SDR performance of SpeakerLM across diverse speaker registration conditions and varying numbers of registered speakers.
% \footnote{Demo page: https://sites.google.com/view/speakerlm/overview}.
\end{abstract}

\begin{links}
    \link{Code}{https://sites.google.com/view/speakerlm/overview}
    % \link{Datasets}{https://aaai.org/example/datasets}
    % \link{Extended version}{https://aaai.org/example/extended-version}
\end{links}

\section{Introduction}
Speaker Diarization (SD) partitions an audio into homogeneous segments according to speaker identities, effectively answering ``who spoke when''~\cite{sd1_2006,sd2_2012,sd3_2022}.
In real-world scenarios such as multi-speaker meetings, knowing ``who spoke when'' is insufficient. 
Associating speaker labels with speech transcripts is more meaningful, as it addresses the more comprehensive question of ``who spoke when and what''~\cite{misp2025}. We refer to this task as \textbf{Speaker Diarization and Recognition (SDR)}, which jointly performs SD and Automatic Speech Transcription (ASR) for enriched conversational understanding.

Conventional SDR systems typically pipeline an SD module and an ASR module~\cite{sd+asr2_2021,sd_asr3_2022,sd+asr1_2024,sd_asr4_2024}. The SD module segments the audio stream into speaker-homogeneous regions, which are then aligned with the ASR outputs, and the speech content for each identified speaker is transcribed. However, such cascaded systems suffer from several limitations. First, error propagation is a major concern. Inaccuracies in SD, such as incorrect speaker boundaries or label assignments, are passed on to the ASR module, leading to degraded transcription quality and incorrect speaker-attributed text~\cite{sd_asr5_2023}. In addition, conventional SD systems struggle to handle overlapping speech, which is common in real-world conversations, resulting in limited performance \cite{sd4_2025}. Furthermore, the cascaded system lacks joint optimization as ASR and SD modules are usually trained independently with different datasets and frameworks. The disjoint training paradigm prevents the SDR system from fully leveraging the synergy between SD and ASR tasks, limiting the overall SDR performance.

Two main paradigms have been attempted to address these limitations of cascaded SDR systems. One paradigm, namely Speaker-Attributed ASR (SA-ASR), jointly trains SD and ASR. As a representative end-to-end SA-ASR system, SA-Transformer \cite{alimeeting_2023} incorporates an ASR encoder and a speaker encoder, and achieves competitive SDR results. However, SA-ASR systems can only operate under the condition that pre-extracted speaker embeddings are available.
Another paradigm uses Large Language Models (LLMs)~\cite{llm2_2022, llm1_2024} for post-processing. Due to the disjoint training issue in cascaded SDR systems,  temporal mismatches between the two modules may lead to word-level diarization errors~\cite{diarizationlm_2024}. Prior works~\cite{sd_llm_3_2024,sd_llm_2_2025} explore the strong generalization and reasoning abilities of LLMs to post-process the outputs of a cascaded SDR system, refining speaker attributions and correcting transcription inconsistencies. While such post-hoc adjustments may improve the overall SDR performance, they are inherently limited by the quality of the initial SD and ASR outputs, and cannot fully resolve issues arising from upstream error propagation or misalignment between modalities. Therefore, how to leverage LLMs not just for post-processing but as core components in end-to-end SDR systems remains an open challenge.

To address the limitations of all these prior works, we introduce \textbf{SpeakerLM}, a multimodal LLM (MLLM) specially designed for end-to-end SDR. Furthermore, to better handle real-world complex multi-speaker conversational scenarios, we incorporate a flexible speaker registration mechanism into SpeakerLM, enabling SpeakerLM to versatilely adapt to varying levels of speaker information availability in real-world applications. We propose three different speaker registration conditions, including cases where prior speaker identities are unknown, known, and redundant. 

Our contributions can be summarized as follows:

\begin{itemize}[leftmargin=*,noitemsep]
    \item \textbf{The first MLLM for end-to-end SDR.} To the best of our knowledge, our SpeakerLM is the first MLLM designed for end-to-end SDR. Specifically, we apply an audio encoder and two projectors as the front-end, leading to an encoder-projector-LLM architecture tailored for SDR.
    \item \textbf{A flexible speaker registration mechanism for versatile SDR.} We incorporate a flexible speaker registration mechanism into SpeakerLM, where the prior speaker embeddings are projected and concatenated with audio and text tokens, enabling the model to handle diverse multi-speaker scenarios in real-world applications.
    \item \textbf{Extensive analysis.} To evaluate the effectiveness of SpeakerLM, we build various cascaded SDR baselines.
    % with open-source SD and ASR models that achieve SOTA performance on different benchmarks.
    Results show that SpeakerLM demonstrates strong data scaling capability, outperforming all the baselines on both in-domain and out-of-domain public benchmarks.
    % \footnote{Demo page: https://sites.google.com/view/speakerlm/overview}.
\end{itemize}

\section{Related Work}

\subsection{Speaker Diarization}
Traditional speaker diarization frameworks typically consist of multiple components, including Voice Activity Detection (VAD) to identify speech segments, speaker embedding extraction to represent speaker characteristics, and clustering to group speaker embeddings for identifying speakers. 3D-Speaker \cite{3d_speaker_2025} is a representative traditional SD system following this pipeline, and achieves competitive performance across multiple public benchmarks.

Conventional SD systems struggle with overlapping speech, because VAD is designed under the assumption of single-speaker segments. 
One promising solution to address this issue is to replace the VAD module with a neural speaker activity detection module, which predicts active speakers in each time frame~\cite{e2e_sd_2019,e2e_vc_sd_1_2021,e2e_vc_sd_2_2021}. Along this direction, Pyannote~\cite{pyannote_1_2020,pyannote_2_2023} achieves strong performance on different datasets. Based on Pyannote, Diarizen \cite{diarizen_1_2025,diarizen_2_2025} incorporates representations from the pretrained WavLM model \cite{wavlm_2022} into the neural speaker activity detection module, outperforming Pyannote on various datasets by providing richer acoustic information.

\begin{figure}[t]
\centering
\includegraphics[width=0.82\columnwidth]{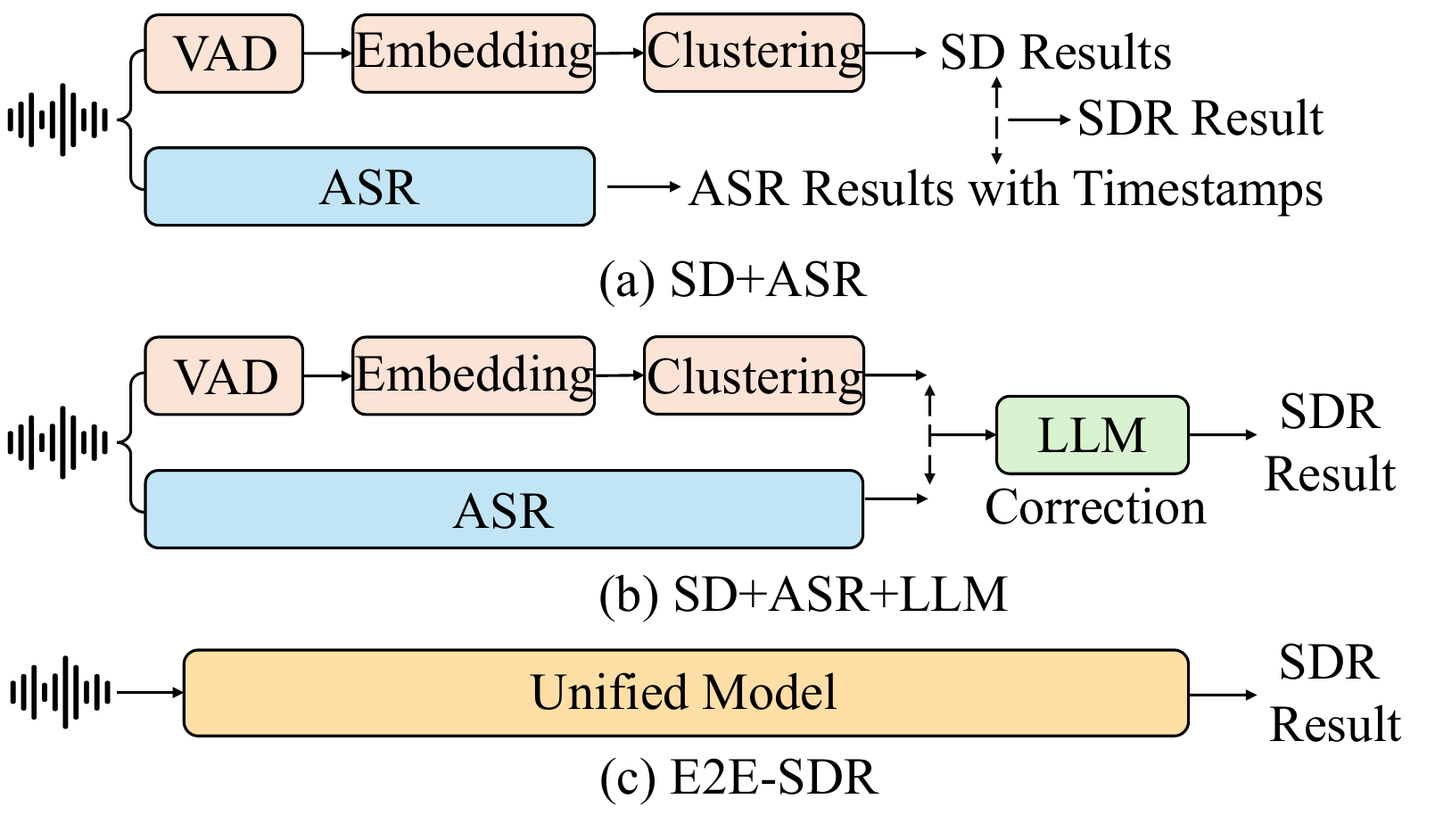}
\caption{Comparison between three different SDR frameworks: (a) SD+ASR (b) SD+ASR+LLM and (c) E2E-SDR.}
\label{fi2:frameworks}
\end{figure}

\subsection{Speaker Diarization and Recognition}
% Figure~\ref{fi2:frameworks} compares three system architectures for SDR. 
Figure~\ref{fi2:frameworks} (a) depicts a conventional SDR system that pipelines SD and ASR modules~\cite{sd+asr1_2024}, denoted as \textbf{SD+ASR}. This pipeline has several limitations, such as error propagation and lack of joint optimization. 
To address these issues, one direction uses LLMs for post-processing, and another direction jointly models SD and ASR tasks within a single unified model in an end-to-end manner. As shown in Figure~\ref{fi2:frameworks}, we denote these two methods by \textbf{SD+ASR+LLM} and \textbf{E2E-SDR}, respectively.

For E2E-SDR, existing studies typically incorporate pre-extracted speaker embeddings for all speakers. This framework, also referred to as SA-ASR \cite{alimeeting_2023}, assumes that the provided speaker embeddings exactly match the speakers in the ground truth. Consequently, it struggles to handle mismatched scenarios such as unregistered speakers or over-registered cases. Our work follows the E2E-SDR paradigm, but is different from the prior SA-ASR systems. Specifically, we introduce a more flexible speaker registration mechanism during training, enabling the model to handle speaker registration inconsistencies effectively. Compared with the cascaded frameworks, our unified modeling approach allows for exploring the synergy between SD and ASR tasks and joint optimization, leading to better alignment between speaker and textual information.

\subsection{Multimodal Large Language Model}
Recently, integrating audio and text with MLLMs has led to notable advances in auditory comprehension and reasoning. Representative models, such as MinMo \cite{minmo_2025}, Qwen2-Audio \cite{qwen2-audio_2024} and Kimi-Audio \cite{kimi-audio_2025}, have demonstrated strong capabilities in audio-text understanding and generation tasks. However, most existing MLLMs primarily focus on single-speaker scenarios, where they exhibit strong performance in ASR tasks but leave SD tasks unexplored. More recently, the Multi-Talker LLM (MT-LLM) has been proposed to handle multi-speaker ASR \cite{mt_llm_2025}. However, MT-LLM is limited to detecting speaker change points and do not explicitly tackle speaker attribution within multi-speaker transcripts.

Based on our review of the literature, SpeakerLM is the first audio-text MLLM capable of performing SDR, integrating both speech content understanding and speaker-aware modeling within a unified framework. Unlike prior works, SpeakerLM goes beyond basic ASR or change-point detection and explicitly explores the performance of speaker assignment in multi-speaker transcription tasks.

\begin{figure*}[t]
\centering
\includegraphics[width=1.65\columnwidth]{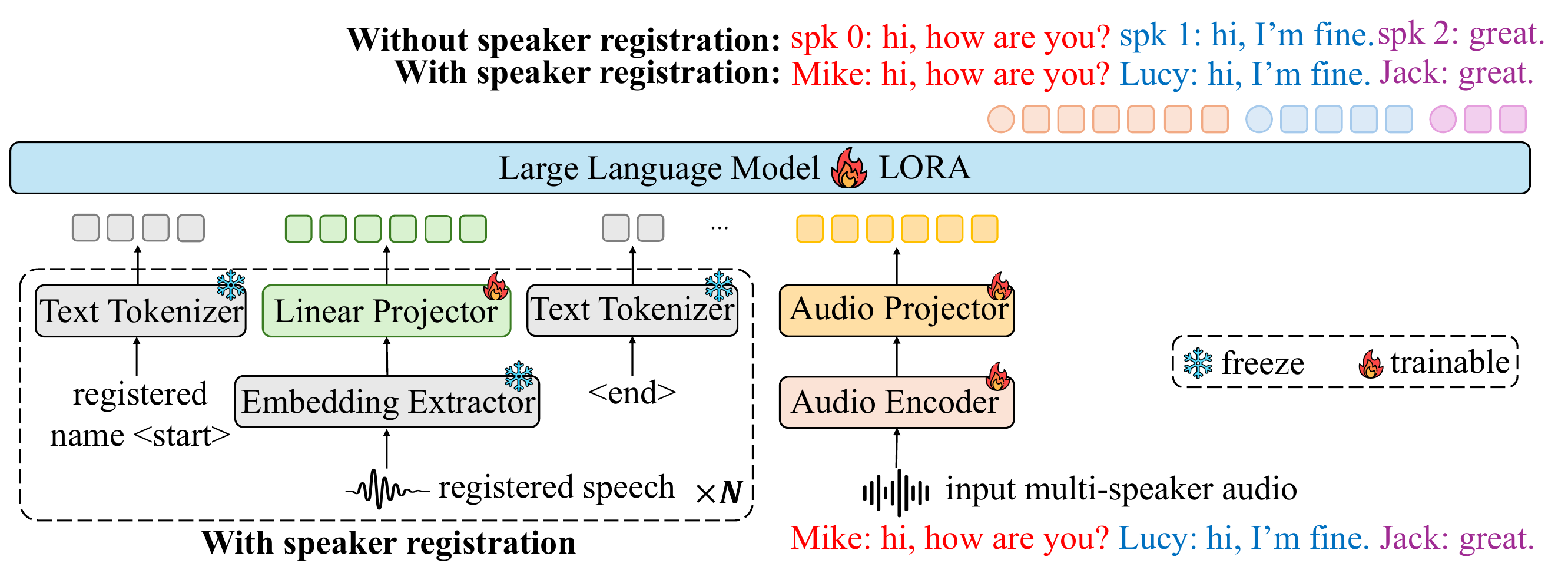}
\caption{Overall architecture of the proposed \textbf{SpeakerLM}. When speaker registration is not performed, each speaker in the transcript is identified by an anonymous ID. With speaker registration enabled, speakers are labeled by their actual names.}
\label{fi3:speakerlm}
\end{figure*}

\begin{figure}[t]
\centering
\includegraphics[width=1\columnwidth]{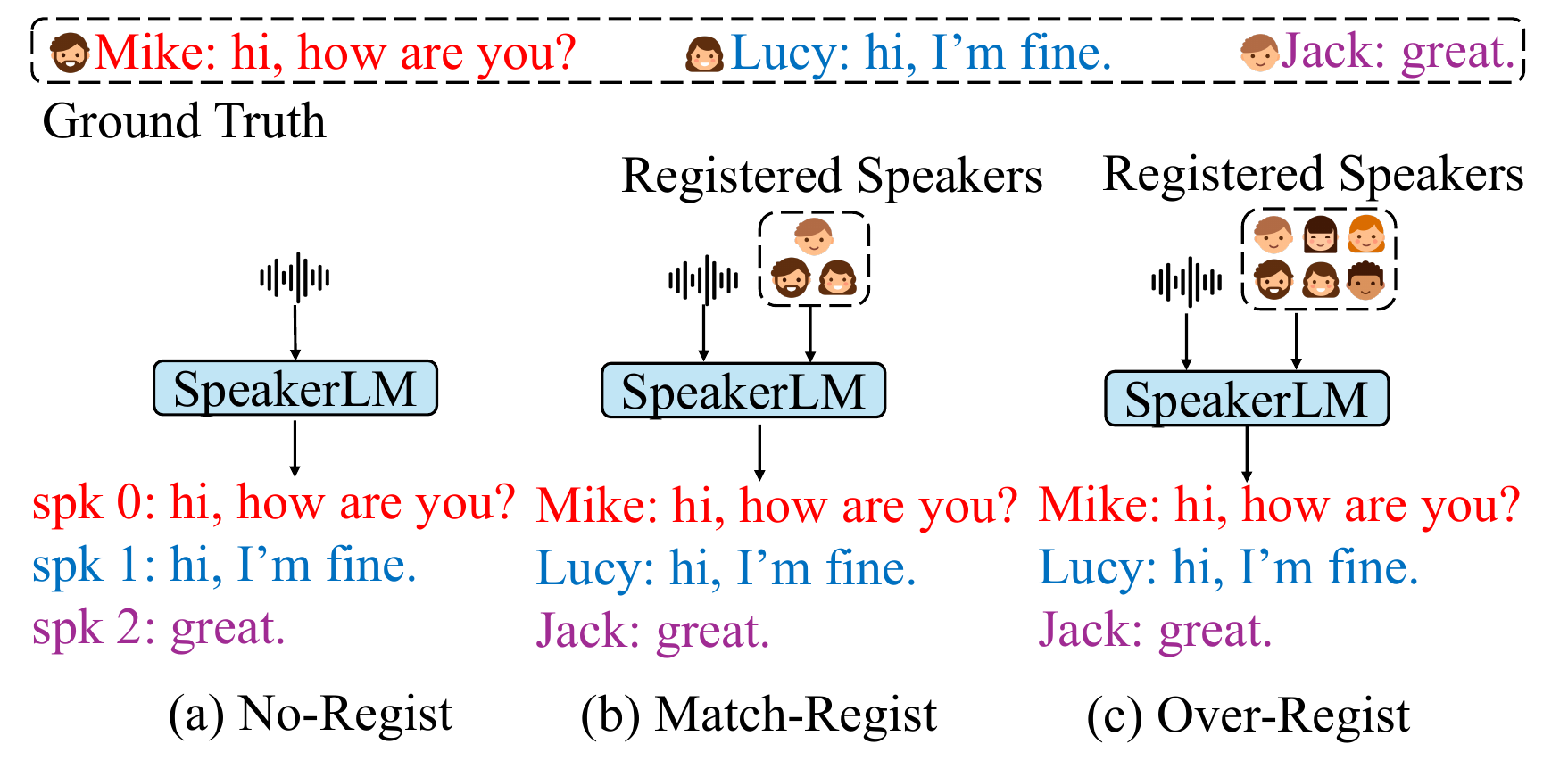}
\caption{Overview of the proposed three different speaker registration mechanisms: (a) No-Regist (no speaker registration) (b) Match-Regist (exact actual speakers in the audio are pre-registered) (c) Over-Regist (more speakers are registered than actual speakers in the audio). For Match- and Over-Regist, speakers are registered in a random order.}
\label{fi4:regist}
\end{figure}

\section{SpeakerLM}
\label{sec:speakerlm}

\subsection{Model Architecture}
Figure~\ref{fi3:speakerlm} illustrates the overall architecture of SpeakerLM.
Specifically, SpeakerLM integrates a 
\textit{lightweight} modality alignment mechanism into a pre-trained text LLM.
For input multi-speaker audio, we use an audio encoder for encoding, followed by a projector to inject the audio embedddings into the feature space of the text LLM.
For speaker registration, a frozen text tokenizer is used to tokenize both the registered speaker's name and the special markers (i.e., \texttt{<start>} and \texttt{<end>}).
The registered speaker's speech is first processed by a frozen pre-trained embedding extractor to obtain speaker embeddings, which are then projected by a single linear layer into the LLM backbone.

\textbf{Audio Encoder and Projector.}
The audio encoder is initialized with the pre-trained SenseVoice-large encoder \cite{sensevoice_2024}, which offers robust audio representation capabilities and provides strong performance across various audio understanding tasks such as multilingual speech recognition and audio event detection.
For the audio projector, we use a randomly initialized two-layer Transformer \cite{transformer_2017}, followed by a convolutional neural network layer for dimensional alignment.

\textbf{Embedding Extractor and Projector.}
A pre-trained speaker embedding model is utilized to extract speaker embeddings, delivering robust and discriminative representations that are essential for precise speaker identification and attribution. Specifically, we use an open-source speaker embedding model ERes2NetV2 \cite{eres2netv2_2024} for embedding extraction, which has achieved SOTA performance on multiple speaker verification benchmarks. A single linear layer projector is applied to the extracted embedding for dimensional alignment.

\textbf{Large Language Model.} 
We use the pre-trained Qwen2.5-7B-Instruct model \cite{qwen2.5_2024} as the backbone text LLM to benefit from its strong instruction-following and general language understanding capabilities, which allows SpeakerLM to effectively handle complex multi-speaker SDR tasks with different levels of information.

\subsection{Flexible Speaker Registration Mechanism}
As shown in Figure \ref{fi3:speakerlm}, we incorporate a flexible speaker registration mechanism into SpeakerLM.
In order to facilitate real-world scenarios, we propose three different registration strategies, including No-Regist, Match-Regist, Over-Regist, as illustrated in Figure \ref{fi4:regist}.
Given the number of speakers in the ground truth as $N_{gt}$ and the number of registered speakers as $N_{rg}$, the relationship between these two values in different registration settings is formulated as follows:
\begin{equation}
\label{eq:regist}
     N_{rg} = 
        \begin{cases}
        0, & \text{if No-Regist} \\
        N_{gt}, & \text{if Match-Regist} \\
        N_{gt} + N_{ov}, & \text{if Over-Regist}
        \end{cases}
\end{equation}
where $N_{ov}>0$ is the number of over-registered speakers.

\textbf{No-Regist} means no speaker registration is performed. This is the conventional setting for cascade SD systems and applications.
We simply feed multi-speaker audio into the model without providing any prior information about the speakers. 
This paradigm is align with the traditional cascade SDR framework, where each speaker in the output is represented by a anonymous speaker ID (e.g., spk 0, spk 1, ...).

\textbf{Match-Regist} assumes that every speaker appearing in the ground truth has been registered in advance, same as SA-ASR. 
The model is required to correctly assign each speaker to the corresponding name. 
This setting closely reflects scenarios where known users are pre-registered, and personalized outputs (e.g., transcripts with speaker names) are desirable. 
For Match-Regist, accurate speaker-name association becomes essential, and the model must effectively leverage the registered speaker information for identification.

\textbf{Over-Regist} refers to the case where more speakers are registered than actually appear in the audio. 
In this scenario, the model must determine which registered speakers are not present in the current utterance, and correctly perform speaker-attributed SDR for the remaining active speakers. 
This setup is more challenging than Match-Regist, as the model must handle redundant speaker information and suppress irrelevant identities. 
It also aligns well with practical scenarios, where a large pool of users may be pre-registered, but only a subset participate in a given interaction.

Overall, the proposed speaker registration framework allows SpeakerLM to flexibly perform SDR under different speaker supervision levels, making it applicable to various practical multi-speaker tasks, from anonymous transcription to personalized speaker transcription.

\subsection{Multi-Stage Training Strategy}
\label{sec:multi-stage}

We apply a four-stage training strategy to progressively enhance the model’s audio understanding capabilities.

\textbf{Stage 1.} To enhance the ASR performance, we first train SpeakerLM using only ASR data, leading to an ASR model, denoted by \textbf{SpeakerLM-ASR}.
To efficiently adapt the LLM while preserving its pre-trained language capabilities, we apply a Low-Rank Adaptation (LoRA) \cite{lora_2022} strategy. Following the ASR training setup of MinMo \cite{minmo_2025}, we sample 600,000 hours of audio from publicly available ASR datasets for training. In this stage, the speaker embedding extractor and projector are not included in the model. In the subsequent three stages, these modules are incorporated into the whole architecture.

\textbf{Stage 2.} We train the randomly initialized projectors using simulated SDR data while keeping both the LLM and audio encoder frozen, which aims to rapidly align the audio and text for SDR domain. Training on simulated data allows the projectors to establish an initial coarse alignment under a simplified distribution. Compared with real recordings, the simulated mixtures are created by simply concatenating utterances from different speakers, without modeling very strong noise or reverberation.

\textbf{Stage 3.} In the remaining two stages, we use real SDR data for training, to further adapt the model to practical acoustic conditions and speaker diversity. In this stage, we jointly fine-tune the audio encoder and the projectors, while keeping the LLM frozen, which allows the model to better capture acoustic patterns and characteristics.

\textbf{Stage 4.} Finally, we jointly fine-tune all modules while applying LoRA to the LLM. This stage enables jointly optimization to better integrate linguistic and acoustic information in complex multi-speaker SDR tasks.

\section{Data Composition}

\subsection{Real Data Collection}
In this work, we conduct experiments on Mandarin SDR datasets. Overall, we sample 238.55 hours of audio clips from public corpora for training and evaluation, covering various real-world multi-speaker scenarios. In addition, we use 7456.99 hours of in-house data for training and validation, to further enhance the model performance. Detailed statistics are summarized in Table \ref{tab:real_data}.

Specifically, the public Alimeeting \cite{alimeeting_2023} and AISHELL4 \cite{aishell4_2021} datasets are collections of multi-speaker conversation recordings in meeting rooms, including both far-field overlapping audio clips and near-field recordings for each speaker. Following previous works \cite{diarizen_1_2025,3d_speaker_2025}, we use the first channel of the far-field signals for training and evaluation. The in-house data contains multi-speaker audio samples recorded in professional studios. The AISHELL5 dataset \cite{aishell5_2025} was recorded using far-field microphones in a hybrid electric car. The dataset includes diverse external noise sources such as wind and tire, along with internal noises from music and air conditioning, making it a more challenging setting than standard meeting room or studio environments. The dataset has two official test set splits for ASR and SDR tasks respectively. We use the first channel of the far-field audio from the SDR test split as the test set in this work, denoted by AISHELL5-Eval.
Notably, AISHELL5-Eval presents a challenging acoustic condition that is significantly different from the domains of the training data, hence it is primarily used to evaluate the out-of-domain SDR performance (\textbf{generalizability}) of the proposed SpeakerLM.

\begin{table}[htbp]
\centering
\renewcommand\arraystretch{0.8}{
\setlength{\tabcolsep}{2.5mm}{
\begin{tabular}{cccc}
\toprule
Dataset & Duration (h) & \# Spk & Usage \\
\midrule
AliMeeting-Train & 104.75 & 2$\sim$4 & Train \\
AISHELL4-Train & 107.50 & 4$\sim$8 & Train \\
In-House-Train & 7,426.70 & 2$\sim$7 & Train \\
In-House-Valid & 30.29 & 2$\sim$7 & Valid \\
AliMeeting-Eval & 10.00 & 2$\sim$4 & Test \\
AISHELL4-Eval & 12.72 & 5$\sim$7 & Test \\
AISHELL5-Eval & 3.58 & 2 & Test \\
\bottomrule
\end{tabular}
}
}
\caption{Statistics of the real SDR data, where ``\#Spk'' denotes the number of speakers in each recording.}
\label{tab:real_data}
\end{table}

\subsection{Simulated Data Creation}
As described in Section~\ref{sec:multi-stage}, we apply a four-stage training strategy to train SpeakerLM.
In the second stage, we use simulated data for quickly training the projectors and audio-text alignment.
Specifically, we use the near-field speech samples from AliMeeting-Train, AISHELL2~\cite{aishell2_2018}, Librispeech~\cite{librispeech_2015} and In-House-Train for SDR data simulation.
The near-field speech samples are randomly mixed to simulate multi-speaker conversation segments,
with each segment lasting 50 seconds and containing 2 to 4 speakers. 
Following previous work~\cite{simulated_2022}, we randomly add real-world noises and reverberations to the clean speech samples, with signal-to-noise ratios (SNRs) uniformly sampled from [10dB, 20dB].
In total, the simulated data comprise 5,000 hours audio samples for training and 5.6 hours for evaluation; the test split is denoted by Simulation-Test.

\section{Experiments}

\subsection{Implementation Details}
The majority of audio samples are recorded at 16 kHz and any samples with a different sampling rate are resampled to 16 kHz to ensure uniformity.
Recordings are randomly split into clips ranging from 40 to 50 seconds in duration to train and test SpeakerLM. For speaker registration, the speech of the registered speaker is segmented into 2-10 seconds clips for embedding extraction, and the corresponding embeddings are averaged to generate a single representative speaker embedding. For Over-Regist, the number of over-registered speakers (i.e., $N_{ov}$ in Eq.~\ref{eq:regist}) uniformly ranges from 1 to 50 during training.

We use the AdamW~\cite{adamw_2024} optimizer for optimization. The learning rate is initially set to 1e-5, linearly increased to 5e-5 during the warm-up phase, and then gradually decayed following a cosine schedule. We use a dynamic batching strategy, adjusting the number of samples per batch based on their length, with a maximum token limit of 6K to balance efficiency and memory usage. Models are trained on 4 NVIDIA A800 GPUs for 1M steps at each stage, with a validation step of 10K. 
The training pseudocode and the specific prompts used for LLMs are presented in code page.

\begin{table*}[htbp]
\centering
\renewcommand\arraystretch{0.85}{
\setlength{\tabcolsep}{1.1mm}{
\begin{tabular}{c|c|c|ccc|ccc|ccc}
\toprule
\multirow{2}{*}{\textbf{System}} & \multirow{2}{*}{\textbf{Params}} & \multirow{2}{*}{\textbf{\#Models}} & \multicolumn{3}{c|}{\textbf{Alimeeting-Eval}} & \multicolumn{3}{c|}{\textbf{AISHELL4-Eval}} & \multicolumn{3}{c}{\textbf{AISHELL5-Eval}}\\
& & & CER$\downarrow$ & cpCER$\downarrow$ & $\Delta\textrm{cp}\downarrow$ & CER$\downarrow$ & cpCER$\downarrow$ & $\Delta\textrm{cp}\downarrow$ & CER$\downarrow$ & cpCER$\downarrow$ & $\Delta\textrm{cp}\downarrow$\\
\midrule
\multicolumn{12}{l}{\textit{\textbf{SD+ASR}}} \\
\midrule
3D-Speaker+Para  & 70M & 4 & 21.30 & 24.94 & 3.64 & 23.02 & 26.01 & 2.99 & 60.16 & 64.12 & 3.96\\
Pyannote+Para  &  70M & 4 & 21.30 & 24.45 & 3.15 & 23.02 & 28.22 & 5.20 & 60.16 & 68.37 & 8.21\\
DiariZen-base+Para  & 95M & 4 & 21.30 & 23.97 & 2.67 & 23.02 & 27.27 & 4.25 & 60.16 & 66.89 & 6.73\\
DiariZen-large+Para  & 140M & 4 & 21.30 & 23.20 & 1.90 & 23.02 & 25.78 & 2.76 & 60.16 & 61.81 & 1.65\\
\midrule
\multicolumn{12}{l}{\textit{\textbf{SD+ASR+LLM}}}\\
\midrule
ChatGPT4.5 (z.s.) & - & 5 & 30.63 & 38.64 & 8.01 & 33.02 & 39.21 & 6.19 & 67.34 & 79.05 & 11.71\\
Qwen2.5-7B-Instruct (z.s.)  & 7B & 5 & 40.10 & 51.01 & 10.91 & 33.92 & 44.16 & 10.24 & 65.67 & 73.30 & 7.63\\
Qwen2.5-7B-Instruct (f.t.)  & 7B & 5 & 21.38 & 22.65 & \textbf{1.27} & 23.05 & 24.93 & 1.88 & 60.17 & 61.63 & 1.46\\
\midrule
\multicolumn{12}{l}{\textit{\textbf{E2E-SDR}}} \\
\midrule
SpeakerLM-ASR & 7B & 1 & 20.98 & - & - & 18.02 & - & - & 55.50 & - & -\\
SpeakerLM (212.25h) & 7B & 1& 18.63 & 32.22 & 13.59 & 17.75 & 26.14 & 8.39 & 48.40 & 64.96 & 16.56 \\
SpeakerLM (694.06h) & 7B & 1& 18.14 & 29.60 & 11.46 & 17.48 & 25.28 & 7.80 & 48.39 & 54.81 & 6.42\\
SpeakerLM (2,269.57h) & 7B & 1 & 17.32 & 27.97 & 10.65 & 17.32 & 23.10 & 5.78 & 47.78 & 50.04 & 2.26\\
SpeakerLM (7,638.95h) & 7B & 1 & \textbf{13.97} & \textbf{16.05} & 2.08 & \textbf{17.17} & \textbf{18.37} & \textbf{1.20} & \textbf{47.24} & \textbf{47.81} & \textbf{0.57}\\
\bottomrule
\end{tabular}
}
}
\caption{The performance of our SpeakerLM (with different amounts of training data) and the baselines on the \textit{in-domain} AliMeeting-Eval, AISHELL4-Eval, and the \textit{out-of-domain} AISHELL5-Eval set. The best results are boldfaced. For SD+ASR+LLM, we use DiariZen-large+Para as the SD+ASR front-end, where ``z.s.'' is ``zero shot'' and ``f.t.'' is ``fine-tune''. All SpeakerLM variants are evaluated under the \textit{No-Regist} condition. ``\#Models'' represents the number of models included.}
\label{tab:no-regist-1}
\end{table*}

\subsection{Evaluation Setups}
\textbf{Evaluation Metrics.} We evaluate the SDR performance on the commonly used publicly available benchmarks, namely, the in-domain AliMeeting-Eval and AISHELL4-Eval, and the out-of-domain AISHELL5-Eval.
Following previous works \cite{alimeeting_2023,diarizationlm_2024,aishell5_2025}, we use Character Error Rate (CER), concatenated minimum permutation CER (cpCER), $\Delta \textrm{cp}$, speaker-attributed CER (saCER), and $\Delta \textrm{sa}$ as evaluation metrics.

CER measures ASR performance by comparing the predicted transcripts against the ground-truth text, ignoring speaker identities. In contrast, cpCER and saCER jointly evaluate both ASR and SD, calculated by comparing the predicted speaker-attributed transcripts against the ground-truth transcripts assigned to each speaker, hence reflecting the overall system performance for the SDR task. Specifically, for the no-speaker-registration scenario, cpCER is computed after conducting the speaker alignment between the reference and predicted speaker-attributed transcripts through minimum edit distance, which finds the optimal permutation of speaker labels to minimize the CER.
For the registered-speaker scenario where speaker identities are known in advance, saCER is computed by directly aligning the predicted and reference transcripts based on speaker names, without performing permutation search. $\Delta \textrm{cp}$ denotes the difference between cpCER and CER:
\begin{equation}
    \Delta \textrm{cp} = \textrm{cpCER} - \textrm{CER}
\end{equation}
$\Delta \textrm{cp}$ reflects the performance drop due to speaker attribution errors, hence serving as a measure of the SD performance.
Similarly, $\Delta \textrm{sa}$ is calculated as:
\begin{equation}
    \Delta \textrm{sa} = \textrm{saCER} - \textrm{CER}
\end{equation}

All metrics are measured in percentage. Lower values represent better performance. 
We use \textbf{cpCER and saCER as the primary metrics} since they provide a comprehensive measure of both SD and ASR performance.

\textbf{Baseline Systems.}
Figure \ref{fi2:frameworks} depicts three different frameworks for SDR, i.e., SD+ASR, SD+ASR+LLM, and E2E-SDR.
The proposed SpeakerLM belongs to the E2E-SDR paradigm, where a unified model is applied for all tasks in an end-to-end manner.
We build four baseline systems for the SD+ASR paradigm using SOTA open-source models, and apply competitive pre-trained LLMs to correct the outputs from SD+ASR systems, leading to three different SD+ASR+LLM baselines.

For the SD+ASR paradigm, we use Paraformer-large \cite{paraformer_2022} as the ASR module, which was pre-trained on 60K hours of audio data and achieves SOTA performance on various Mandarin ASR benchmarks. For the SD module, we apply four SOTA open-source SD toolkits, including 3D-Speaker \cite{3d_speaker_2025}, Pyannote 3.1 \cite{pyannote_2_2023}, Diarizen-base \cite{diarizen_1_2025}, and Diarizen-large \cite{diarizen_2_2025}. Among the four SD models, 3D-Speaker and Diarizen-large have achieved SOTA performance on different benchmarks, while Pyannote 3.1 and Diarizen-base demonstrate competitive performance. By combining the four SD systems with Paraformer-large, we obtain four SD+ASR baseline systems, namely, \textbf{3D-Speaker+Para}, \textbf{Pyannote+Para}, \textbf{Diarizen-base+Para}, and \textbf{Diarizen-large+Para}.

SD+ASR+LLM refers to a pipeline where a text LLM is used to correct the outputs from the SD+ASR system. For SD+ASR+LLM baselines, prior works using a text LLM for SD+ASR correction are mainly based on English datasets \cite{diarizationlm_2024,sd_llm_3_2024,sd_llm_2_2025}. Following the paradigm, we adapt Qwen2.5-7B-Instruct \cite{qwen2.5_2024} and ChatGPT4.5 \cite{gpt4.5_2025} for correction. We first evaluate their zero-shot performance (z.s.) using prompts following prior work~\cite{sd_llm_2_2025}, then fine-tune Qwen2.5-7B-Instruct (f.t.) on our training datasets for SD+ASR correction.

\subsection{Performance without Speaker Registration}
We first evaluate the performance of SpeakerLM without speaker registration, where each speaker in the prediction results is represented by an anonymous ID. 

\begin{figure}[t]
\centering
\includegraphics[width=1\columnwidth]{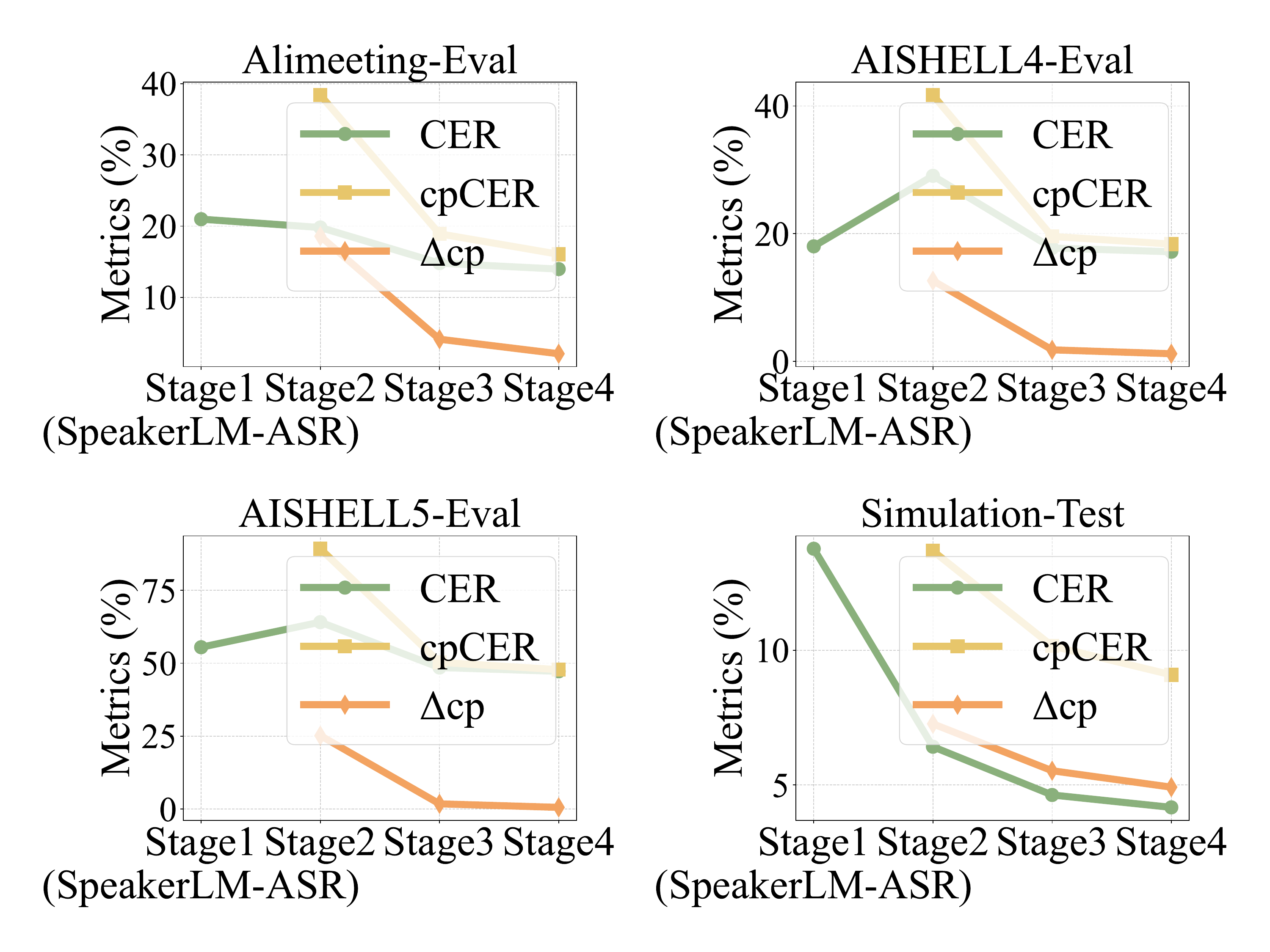}
\caption{The performance of SpeakerLM on test sets under \textit{No-Regist} condition across different training stages.}
\label{fi5:stage}
\end{figure}

\textbf{Comparison with Baselines.} Table~\ref{tab:no-regist-1} compares our SpeakerLM and the baselines under \textbf{No-Regist} condition on various test sets. For SD+ASR with four models included, Diarizen-large+Para achieves the best performance. Compared to 3D-Speaker, both Pyannote and Diarizen employ a neural speaker activity prediction model to replace the VAD module, enabling more effective handling of overlapping speech. Building upon the Pyannote pipeline, Diarizen series further enhances the SD performance by incorporating acoustic features from the pre-trained WavLM model.

For SD+ASR+LLM baselines pipelined with five models, we incorporate a pre-trained text LLM to correct the mis-assigned speaker labels in the results from the best SD+ASR system (i.e., Diarizen-large+Para). Results show that both Qwen2.5-7B-Instruct and ChatGPT-4.5 perform poorly in zero-shot SDR correction, even degrading the overall performance. This finding is consistent with observations in prior works \cite{sd_llm_3_2024, sd_llm_2_2025}. This is mainly due to hallucination of LLMs: despite that we explicitly instruct the LLM to only modify speaker labels, the LLM often alters the content of the speakers’ utterances, leading to a significant increase in CER. Finetuning Qwen2.5-7B-Instruct substantially mitigates the hallucination problem, leading to slight SDR performance improvement over Diarizen-large+Para.

Different from the cascaded architectures of SD+ASR and SD+ASR+LLM, the proposed SpeakerLM is a unified E2E-SDR model that jointly performs speaker diarization and recognition. To explore how much SpeakerLM benefits from the ASR training stage, we first report the CER of SpeakerLM-ASR, which is trained on the ASR data mentioned in Section~\ref{sec:multi-stage}. Compared with Paraformer-large, SpeakerLM-ASR is trained on a larger ASR dataset and thus achieves lower CER across all test sets, which naturally contributes to a lower cpCER for SpeakerLM. However, it should be noted that $\Delta\textrm{cp}$ is not affected by the model’s ASR performance, which serves as a reliable metric for evaluating the speaker attribution performance.

Initializing SpeakerLM by SpeakerLM-ASR, we conduct the following training with different scales of SDR data, ranging from 212.25 hours (Alimeeting-Train+AISHELL4-Train) to 7,638.95 hours. Results show that when trained on a limited amount of SDR data, SpeakerLM falls behind most cascaded baselines. However, as the training data scale up, SpeakerLM demonstrates a strong data scaling capability, with substantial improvements in cpCER and $\Delta\textrm{cp}$. Moreover, we observe that the improvements on CER are marginal. This is because the in-house data mainly consists of near-field recordings, which offer limited ASR gains for far-field speech with reverberation. With 7,638.95 hours training data,  SpeakerLM significantly outperforms all baselines. In terms of cpCER, SpeakerLM achieves absolute improvements of 6.60\%, 6.56\%, and 13.82\% over the strongest cascaded system on AliMeeting-Eval, AISHELL-4-Eval, and AISHELL-5-Eval, respectively. Notably, SpeakerLM also achieves superior performance on the out-of-domain and challenging test set, AISHELL5-Eval, with a $\Delta\textrm{cp}$ of \textbf{0.57}, indicating \textbf{strong robustness and generalizability of SpeakerLM to unseen and noisy acoustic environments}.

\textbf{Impact of Multi-Stage Training.} % 图片
Figure~\ref{fi5:stage} illustrates the performance of SpeakerLM across the proposed four different training stages (Section~\ref{sec:multi-stage}).
On both Alimeeting-Eval and Simulation-Test, the performance of SpeakerLM improves progressively, demonstrating the effectiveness of the proposed multi-stage training strategy. However, on AISHELL4-Eval and AISHELL5-Eval, CERs at Stage 2 are higher than those at Stage 1. This is because Stage 2 relies on simulated data, while the simulation process does not incorporate any audio data from these two datasets, leading to a domain mismatch. With four training stages, SpeakerLM demonstrates strong generalizability on the out-of-domain test set, i.e, AISHELL5-Eval. This suggests that the subsequent training stages (Stage 3 and Stage 4), which involve fine-tuning on more realistic and diverse meeting-style data, are crucial for mitigating domain mismatch and enhancing the model’s robustness across different evaluation scenarios. In addition, the ASR performance on simulated data is consistently better than that on real data, which is attributed to the simplicity of the simulated data. Nevertheless, we believe the simulation process remains meaningful, as it closely resembles office or workplace scenarios where a voice assistant is used to record meetings.

\begin{figure}[t]
\centering
\includegraphics[width=0.98\columnwidth]{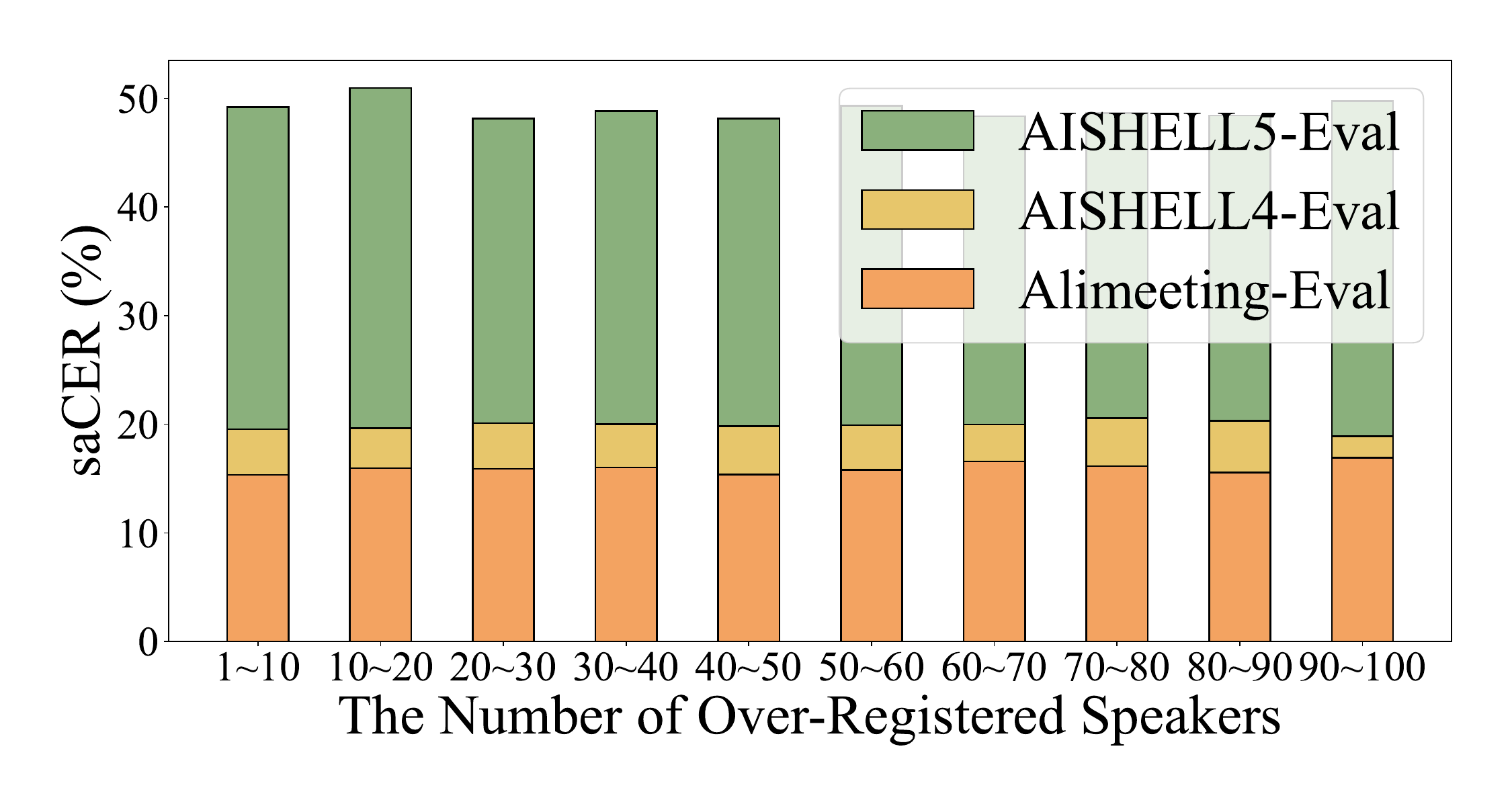} % Reduce the figure size so that it is slightly narrower than the column. Don't use precise values for figure width.This setup will avoid overfull boxes.
\caption{The saCER of SpeakerLM under Over-Regist condition with different numbers of over-registered speakers.}
\label{fi6:sacer}
\end{figure}

\subsection{Performance with Speaker Registration}
In this section, we explore the SDR performance of SpeakerLM with different numbers of registered speakers.

\textbf{Match-Regist vs Over-Regist.}
In Table~\ref{tab:regist-1}, we first present the performance of SpeakerLM and SA-Transformer on AliMeeting-Eval.
Results show that SpeakerLM outperforms SA-Transformer, achieving a 25.98\% absolute improvement in saCER.
Then, we compare the performance of SpeakerLM under different conditions.
For Over-Regist, $N_{ov}$ randomly ranges from 1 to 50, consistent with the training setup.
In terms of CER and cpCER, SpeakerLM shows comparable performance across the two registration mechanisms. However, for $\Delta{\textrm{sa}}$, the metrics are consistently lower under the Match-Regist setting, suggesting that the presence of redundant registered speakers introduces unnecessary information, negatively affecting the SD performance.

\begin{table}[htbp]
\centering
\renewcommand\arraystretch{0.6}{
\setlength{\tabcolsep}{0.3mm}{
\begin{tabular}{c|c|ccc}
\toprule
Dataset & System (RM) & CER & saCER & $\Delta \textrm{sa}$ \\
\midrule
\multirow{3}{*}{Alimeeting-Eval} & SA-Transformer (MR){$^{*}$} & - & 41.55 & - \\
& SpeakerLM (MR) & 13.98 & \textbf{15.57} & \textbf{1.59} \\
 & SpeakerLM (OR) & \textbf{13.96} & 15.71 & 1.75 \\
 \midrule
 \multirow{2}{*}{AISHELL4-Eval} & SpeakerLM (MR) & \textbf{17.13} & \textbf{19.73} & \textbf{2.60}\\
 & SpeakerLM (OR) & 17.15 & 20.16 & 3.01\\
 \midrule
 \multirow{2}{*}{AISHELL5-Eval} & SpeakerLM (MR) & 47.05 & 47.36 & \textbf{0.31}\\
 & SpeakerLM (OR) & \textbf{46.69} & \textbf{47.35} & 0.66\\
\bottomrule
\end{tabular}
}
}
\begin{flushleft}
\footnotesize
* Due to the lack of an open-source implementation, we adapt the reported results from the original work for comparison.
\end{flushleft}
\caption{The performance of different systems on AliMeeting-Eval, AISHELL4-Eval and AISHELL5-Eval. ``RM'', ``MR'' and ``OV'' represent ``Register Mode'', ``Match-Regist'' and ``Over-Regist'', respectively.}
\label{tab:regist-1}
\end{table}

\textbf{Impact of the Number of Over-Registered Speakers.} % 图片
Figure~\ref{fi6:sacer} explores the impact of the number of over-registered speakers (i.e., $N_{ov}$) on model performance. 
We do not observe significant performance degradation as $N_{ov}$ increases.
This reflects SpeakerLM's robustness to redundant speaker identities and the ability to focus on relevant speaker representations during inference.

\subsection{Impact of Embedding Extractors} % 表格
In the experiments above, we adopt ERes2NetV2 as the speaker embedding extractor. 
To investigate the impact of the speaker embeddings, we replace ERes2NetV2 with another speaker embedding model, i.e., CAM++ \cite{cam++_2023}, for training and test. 
It should be noted that ERes2NetV2 performs better than CAM++ in various speaker verification benchmarks \cite{3d_speaker_2025}.
Results in Table~\ref{tab:ablation-1} show that SpeakerLM using embeddings from ERes2NetV2 outperforms the variant with CAM++, indicating that improvements in the speaker embedding model can further enhance the overall performance of SpeakerLM.

\begin{table}[htbp]
\centering
\renewcommand\arraystretch{0.6}{
\setlength{\tabcolsep}{1.6mm}{
\begin{tabular}{c|c|ccc}
\toprule
Register Mode & Extractor & CER & cpCER & saCER \\
\midrule
\multirow{2}{*}{No-Regist} & CAM++ &  14.63 & 16.74 & -\\
& ERes2NetV2 & \textbf{13.97} & \textbf{16.05} & -\\
\midrule
\multirow{2}{*}{Match-Regist} & CAM++ & 14.74 & - &	17.23 \\
& ERes2NetV2 & \textbf{13.98} & - & \textbf{15.57}\\
\midrule
\multirow{2}{*}{Over-Regist} & CAM++ & 14.71 & - & 16.92 \\
& ERes2NetV2 & \textbf{13.96} & - & \textbf{15.71}\\
\bottomrule
\end{tabular}
}
}
\caption{The performance of SpeakerLM on AliMeeting-Eval with different embedding extractors.}
\label{tab:ablation-1}
\end{table}

\section{Conclusions}
In this work, we present SpeakerLM, the first MLLM designed for end-to-end SDR. We introduce a flexible registration mechanism into SpeakerLM, enabling the model to perform SDR with different numbers of registered speakers. Compared with strong cascaded baselines built from SOTA SD and ASR models, SpeakerLM achieves superior performance on both in-domain and out-of-domain test sets. This work demonstrates the potential of MLLMs to model SDR in a scalable, data-driven, and autoregressive manner.

\bibliography{aaai2026}

\end{document}